# Sub-cm Resolution Distributed Fiber Optic Hydrogen Sensing with Nano-Engineered TiO$_2$

Zsolt Poole[1], Paul Ohodnicki[2], Aidong yan[1], Yuankun Lin[3], and Kevin Chen[1]

[1]*Department of Electrical and Computer Engineering, University of Pittsburgh Pittsburgh, PA, 15261, USA*
[2]*National Energy Technology Laboratory, 626 Cochrans Mill Road, Pittsburgh, PA 15236, USA*
[3]*Department of Physics, University of North Texas, Denton, TX 76203, USA*

**Abstract:** The 3D nano-structuring on the <10nm scale allows the refractive index of functional sensory materials(n>2) to be reduced and matched with the cladding of optical fiber(n~1.46) for low-loss integration. A high temperature capable hydrogen sensor composed of D-shaped optical fiber with palladium nanoparticles infused nanoporous (~5nm) TiO$_2$ film is demonstrated. The behavior of the developed sensor was characterized by examining the wavelength of an incorporated Fiber Bragg Grating and by observing the transmission losses at temperatures up to 700°C. In addition, with frequency domain reflectometry the distributed sensing potential of the developed sensor for hydrogen concentrations of up to 10% is examined. The results show the possibility of detecting chemical gradients with sub-cm resolution at very high temperatures (>500°C).

## 1. Introduction

Optical fiber is a widely used sensor platform due to its small form and its ability to withstand harsh conditions such as high temperatures and corrosive environments[1, 2]. A number of sensing applications for optical fiber already exist[1-5], but the inert nature of silica limits its detection capability. Incorporating sensory materials with optical fiber (e.g. evanescent wave configuration) is a practical way to extend the sensory potential of this platform for the detection of numerous biological and chemical species. A number of advances have already been explored including thin palladium films for hydrogen sensing[4, 6, 7] and films of various oxides and other materials for the detection of a number of chemical species[8-13]. Response types from absorptive based[11] to refractive index based[4, 7-10, 13], to surface plasmon resonance based[5, 6, 14] have been explored.

Metal oxide type chemiresitive sensors are well studied and it is known that a conductometric response is to be expected. It is also known that structuring on the scale that is comparable with the sensory modulated depth (~1-10nm, proportional to the Debye length) can provide significant enhancements in the magnitude and rate of the responses[15-19]. Sensitization with noble metals such as palladium can further enhance the sensory mechanism by influencing the rate of response, the type of the response, and the magnitude of the response[16, 19, 20]. Although, some exploration has already shown that metal oxide thin films can be combined with optical fiber for sensory applications, there exists the need to quantify the functionality of this combination[21, 22] and the need for further advancements. The typical refractive indices of sensory metal oxides are greater than or equal to ~2, whereas the refractive index of silica, typical material of optical fiber, is ~1.46. This large refractive index difference poses difficulties in combining materials with optical fiber.

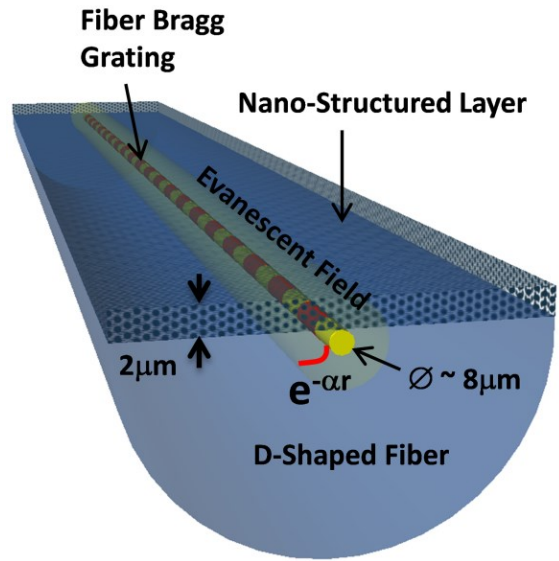

**Fig. 1.** An illustration of the proposed device configuration for evanescent wave based sensing. The Pd nanoparticles infused nanoporous $TiO_2$ interacts with the environment and relays changes to the light in the fiber core through the evanescent field. By including an in-fiber FBG, both absorptive and refractive index changes can be simultenously monitored.

A schematic of the proposed device configuration in provided in **Figure 1** where a Fiber Bragg Grating(FBG) is inscribed into the core and the residual cladding on the flat side of the fiber is removed and replaced with a hydrogen sensitive nanomaterial that interacts with the evanescent field of light propagating in the fiber core. The low-loss integration of $TiO_2$ requires the modification of its refractive index from its nominal value of >2.2, accomplished by controlled 3D structuring on the <10nm scale. In a previous work[21], through finite element modal analysis, the dependence of the film refractive index on its thickness in order to maintain a constant propagation constant in the single mode fiber(constant effective refractive index), was examined. In other words, an approximate mode matching approach is used, maintaining a constant effective refractive index along the fiber after modification, to minimize the incurred losses. It was found that a thick(2μm) nanoporous film with a refractive index close to the refractive index of the fiber cladding should be low-loss when integrated while maximizing the film interaction with the evanescent wave. This finding is only from the optical perspective and does not include the considerable enhancements gained by structuring on the <10nm scale, where a great enhancement in the surface area of the sensory film is obtained.

According to Drude's theory, changes in the free carrier concentration should show up as changes in the absorbed light intensity at near infrared wavelengths[23-25]. The incorporated palladium nanoparticles in the $TiO_2$ matrix[20] are expected to interact with hydrogen to form palladium hydride, resulting in a conductivity and size change[16], which then changes the electronic properties of the film. This, in principle, should result in both light attenuation and a modification of the effective refractive index. Therefore, a Fiber Bragg Grating was integrated into the sensor to allow for the simultaneous observation of changes in optical absorption and changes in the effective refractive index.

Rayleigh backscatter measurements combined with interferometry can be used to locate local variations in the optical fiber with sub-mm resolution. With this technique the distributed sensing potential of the nanoporous hydrogen sensitive film on D-fiber is characterized. It is shown that the demonstrated technique of tuning the refractive index of functional sensory materials for low-loss on-fiber integration can be combined with distributed type measurements for the detection of chemical gradients with sub-mm resolution.

## 2. Experimental

### 2.1 Materials and methods

Methods for the manufacture of the $TiO_2$ nanoporous matrix, using a non-ionic Pluronic type block-copolymer (structure director), were adapted from the literature[26-28]. First, a concentrated Pluronic F-127 solution was prepared by mixing 10g of 1-butanol with 1.3g of 37%HCl and 1.3g of $H_2O$, to which 5g of the block-copolymer was added and dissolved. This solution was subsequently used as the source for the structure directing block-copolymer. Various concentrations of titanium isopropoxide in 1-butanol were prepared to which measured quantities of the polymer source were added to attain a range of the metal source to block-copolymer molar ratios. This ratio determines the final refractive index after processing. The ratio that provided a refractive index close the cladding refractive index of the optical fiber (n~1.46) is the following 1:0.013:1.43:5.64:31.58, composed of titanium isopropoxide, Pluronic F-127, 37%HCl, $H_2O$, and 1-butanol. A separate palladium solution was prepared by dissolving 0.32g of $PdCl_2$ in 6g of 1-butanol and 0.76g of 37% HCl. From this an amount required to obtain a 3mol% ratio to Ti was subsequently added to the prepared Ti-block-copolymer solution. The resultant reddish solution was stirred at room temperature for 2 hours and left to settle for one day before use.

Thin films of the prepared precursor solutions were deposited on ~2x2cm square pieces of <100> silicon wafer by the spin cast method. 100μL of the precursor solutions were deposited after which the samples were quickly accelerated to 2500RPM and held for 30s. After coating, the samples were placed in a furnace and annealed at 400-700°C at a heating/cooling rate of 3°C/minute, and were held for 2 hours at the target temperature. The New Amorphous model[29] was fit to the measured data (parameters Ψ and Δ) to acquire the refractive indices, measured with a Jobin Yvon Spectroscopic Ellipsometer. Variations in the refractive indices were represented by changes in the porosity, modelled using Bruggeman's effective medium theory. The use of effective medium theory is valid since the 3D nanopore dimensions are on the ~λ/40 scale. In air Pd is in the form of PdO nanoparticles in the $TiO_2$ matrix[20] which, due to the small quantity used, did not significantly affect the refractive indices of the measured films. By the presented template based sub-wavelength nano-engineering scheme, nanoporous $TiO_2$ with refractive indices in the range of 1.4-2.2 can be manufactured.

In order to access the fiber core the 4μm cladding material on a 15cm section of the flat side of D-fiber was etched away using buffered HF (21 minutes in $5NH_4F:1HF$). The dip coating method was used to coat the exposed section with the prepared precursor solution, with an approximate rate of ~5mm/s. After applying several coatings, the fiber was left to dry overnight followed by annealing with the procedure already described.

The cross-sectional scanning electron microscope (SEM) image in Figure 2A is an example of a Pd-sensitized TiO$_2$ nanomaterial D-shaped optical fiber sensor, with a film refractive index of ~1.46 and a thickness of ~2μm. Figure 2B is a bright field transmission electron microscope (TEM) image showing the structure of the manufactured sub-wavelength 3D TiO$_2$ matrix. From this image, the dimensions of the nanopores are aproximated to be <10nm.

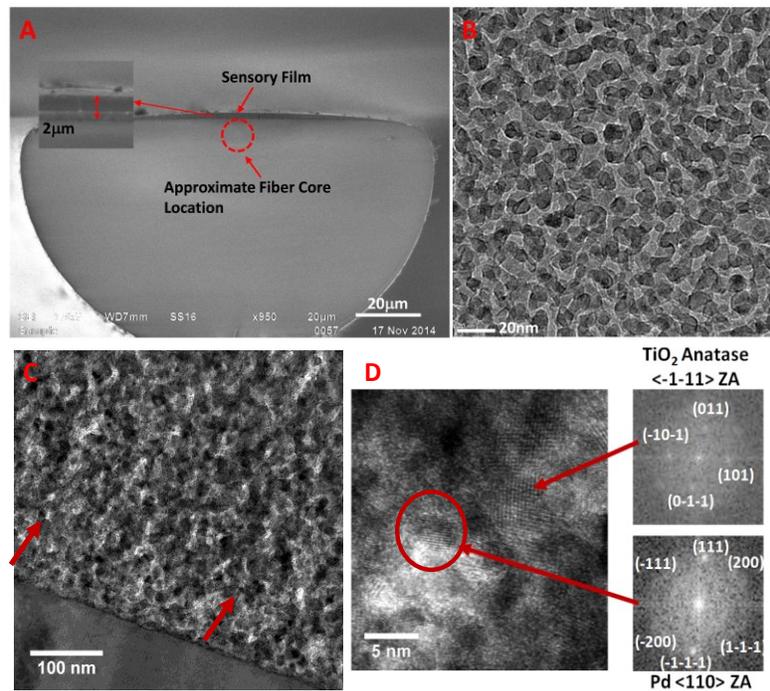

**Fig. 2. (A)** Cross sectional scanning electron microscope (SEM) image of the D-shaped fiber sensor, with a Pd-sensitized nanoporous film thickness of ~2μm and film refractive index of ~1.46. **(B)** Bright Field transmission electron microscope (TEM) image of the TiO$_2$ matrix prepared on SiN$_3$ TEM grids. Nanopore dimensions of <10nm are clearly visible. **(C)** Cross-sectional bright field TEM image at the boundary between the fiber and the porous film, obtained by sectioning with FIB. The arrows indicate dark spots believed to be Pd-based nanoparticles. **(D)** High resolution TEM image for one of the Pd-based nanoparticles with indexed Fast Fourier transform images confirming the presence of Pd.

Cross-sectional transmission electron microscope (TEM) images were taken of one sample that was sectioned with a NOVA focused ion beam (FIB). Clear evidence for Pd-based nanoparticles of approximately 5nm in diameter as well as anatase TiO$_2$ grains can be found by indexing the local diffraction patterns obtained with a Fast Fourier Transform analysis of high-resolution images (Figure 2D). The Pd-based nanoparticles can be observed in the conventional bright field TEM image of Figure 2C as dark spots roughly 5nm in diameter.

*2.2 Sensor characterization*

The dynamics of the sensor were explored from 700°C down to 400°C by cycling between 0.5% hydrogen in nitrogen and pure nitrogen, as shown in **Figure 3A**. Light was coupled into the D-shaped optical fiber from a broadband light source (MPB EBS-7210) with appreciable optical power from 1515nm to 1615nm. The NIR light propagated through the 15cm sensory portion of the fiber and collected on the other side with an InGaAs photo-detector. The hydrogen-induced optical transmission loss, defined as the ratio of the transmitted optical power (P) and without ($P_0$) the presence of hydrogen, is 70% at 700°C. At 400°C the magnitude of the response drops to ~17%. **Figure 3B** shows the sensor response time at 700°C where the transmission loss reached 90% of its steady-state values in 2 minutes and recovered to 90% of its steady-state values in pure nitrogen in 3 minutes. The magnitude of the sensor response is characterized from 200°C to 700°C when exposed to 0.5% hydrogen (**Figure 4B**).

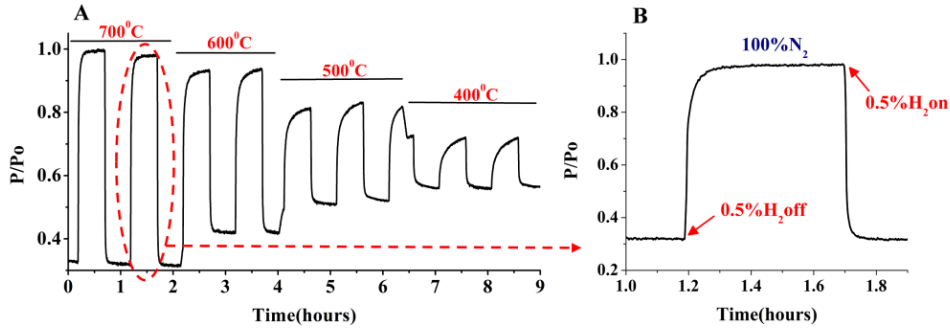

**Fig. 3. (A)** The dynamics of the 3mol% Pd-nanoparticle sensitized nanoporous $TiO_2$ film on D-shaped optical fiber when exposed to 0.5% $H_2$ in nitrogen and recovered with nitrogen, at temperatures of 700°C to 400°C. **(B)** Enlarged view of the selected region at 700°C showing the shape and speed of the response as a function of time.

The introduction of stepwise hydrogen concentration changes from 0.1% to 1% at 600°C and 700°C provided the data shown in **Figure 4A**. The observed strong optical absorption in the presence of a hydrogen partial pressure is likely caused by the formation of palladium hydride providing an increase in the free electron concentration. This, in turn, increases the optical absorption of NIR light in the sensory film. During heat up of the sensor from 25°C to 600°C a factor of 3 increase in the transmitted light power was observed. Therefore, a test was performed to exclude hydrogen's high thermal conductivity being the cause of the observed light intensity variations, that is, the possibility of hydrogen conducting more heat to the sensor as being partly the cause of the sensory response. Helium has a thermal conductivity that is comparably high to that of hydrogen and no response was noted when the sensor was exposed to helium and recovered in nitrogen. One reason behind the temperature dependent increase in the transmitted optical power could be due to an increase in the catalytic absorption of oxygen ions, trapping conduction electrons[15, 30]. This would then show up as an apparent NIR transmission increase due to a reduction in the free electron concentration.

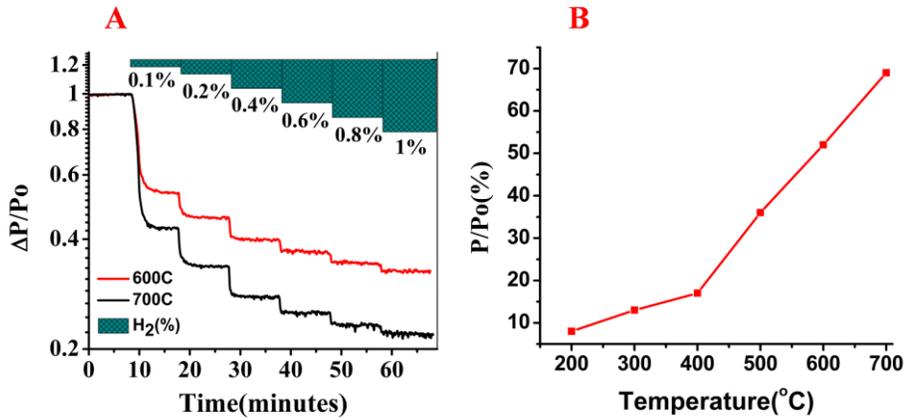

**Fig. 4. (A)** Normalized transmitted optical power (InGaAs photo-detector voltage) as a function of hydrogen concentration at 600°C and 700°C. **(B)** The magnitude of the sensory response as a function of temperature from 200°C to 700°C when exposed to 0.5% hydrogen, balanced with nitrogen.

According to the Kramers-Kronig relation, an absorptive change will be accompanied by a change in the real part of the refractive index, although not necessarily in the same wavelength range. In order to examine whether a measurable change in the real refractive index is present in the Pd-sensitized $TiO_2$ fiber sensor, a fiber Bragg grating was inscribed prior to etching. The fiber was hydrogen loaded for 2 weeks at 1600 psi and a type II fiber Bragg grating (FBG) was inscribed by a phase mask (2.5×1cm with 1060nm period) using a 248nm KrF laser source (GSI Lumonics PM-844) with a cumulative fluence of ~6,000 pulses at ~50mJ/$cm^2$[31, 32]. This was followed by a heating step at 120°C for 24 hours to remove the residual hydrogen.

By using a fiber circulator, both transmitted and reflected light can be simultaneously monitored. A fiber Bragg grating is very sensitive to changes in the environmental refractive indices and will have an associated resonant wavelength shifts($\Delta\lambda_{peak} = 2\Lambda\Delta n_{eff}$). In examining the reflected resonance peak of the fiber Bragg grating when exposed to 2.5% hydrogen, a wavelength displacement was not detected (**Figure 5**). The dimensions of the $TiO_2$ grains are observed to be <10nm with bright field TEM (**Figure 1**), and it is known that the modulated space charge region is ~1-10nm. Therefore, it is reasonable to assume that the $TiO_2$ matrix is fully or near fully modulated when cycled between hydrogen and nitrogen. This is in conjunction with an expected volume increase due to the formation of palladium hydride when Pd nanoparticles are exposed to hydrogen. Therefore, a significant change in the material is expected to take place. The lack of a resonant wavelength shift indicates that there is not a measurable change in the real part of the refractive index of the nanomaterial at the wavelength of interrogation, given the sensitivity of the instruments(OSA resolution ~10pm). This, combined with the confinement factor predicted by the simulation[21] gives a detection limit of ~$5\times10^{-4}$ for observing refractive index variations.

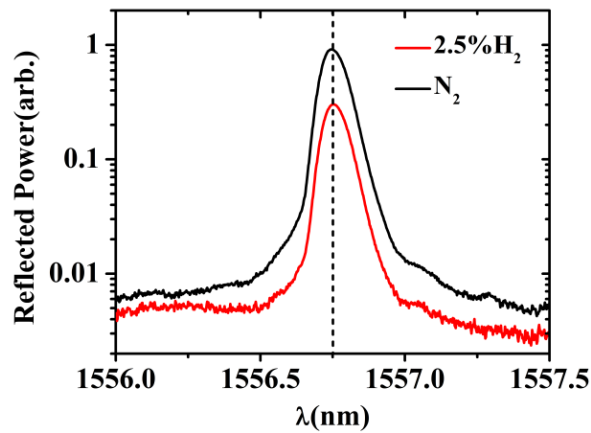

**Fig. 5.** Resonance wavelength of the incorporated Fiber Bragg Grating with a Pd nanoparticle infused $TiO_2$ overlayer before and after exposure to 2.5% hydrogen in nitrogen at 600°C. An FBG wavelength shift was not detected due to hydrogen exposure.

*2.3 Distributed hydrogen sensing*

A distributed type characterization was performed with Luna OBR 4600, using optical frequency domain reflectometry. **Figure 6** highlights the Rayleigh backscatter measurements coupled with a time of flight type analysis(interferometry) for a 5cm section of the constructed sensor as a function of sensor length. The initial data was collected with a spatial resolution of 0.1mm which was subsequently reduced to 1cm, to smooth the data. The irregular features(bumps) in the signal are believed to result from film thickness variations. Given that the film on the fiber was produced by hand drawing through the precursor, thickness variations are expected due to an irregular coating rate.

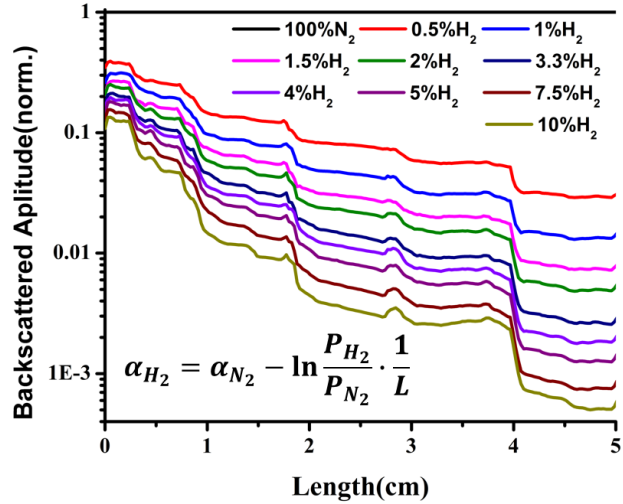

**Figure 6:** Optical frequency domain reflectometry measurement of the backscattered Rayleigh amplitude of a 5cm section of the Pd nanoparticle infused $TiO_2$ D-fiber sensor at 600°C under exposure to various hydrogen concentrations in nitrogen. The backscattered Rayleigh amplitude at different hydrogen concentrations were divided by the Rayleigh amplitude in nitrogen to obtain the data shown, with an approximate equation relating the absorption coefficient in hydrogen to that in nitrogen.

Upon exposure to hydrogen, the modulated free carrier density is accompanied by changes in the absorbed NIR light. The backscattered Rayleigh amplitude is directly proportional to the light intensity at a specific location along the fiber. Therefore, an exponentially decaying propagating light intensity (Beer-Lambert law) due to optical absorption in the sensory film should also generate an exponentially decaying Rayleigh backscatter. The data in **Figure 6** is obtained by dividing the Rayleigh amplitudes in different hydrogen concentrations(balanced with nitrogen) with the amplitude recorded in nitrogen. In this case, an expression for the characteristic attenuation coefficient would take the form shown in **Figure 6**, where P represents the backscattered Rayleigh amplitude at a particular position. On a logarithmic scale straight lines would otherwise be expected indicating the exponential attenuation of light since the current hydrogen exposure scheme does not contain a hydrogen concentration gradient along the sensor length. The observed data provides direct evidence for the possibility of measuring concentration gradients with the developed sensor with a particularly interesting application in hydrogen fuel cells in which gradients are believed to exist due to variations in the fuel consumption across the length of the cell.

### 3. Conclusion

The tuning of the refractive indices of functional sensory materials is demonstrated to have important applications in optical fiber sensing in a prior work[21] and in this paper since with this technique most functional sensory materials can now be optimally integrated with optical fiber. While a Pd nanoparticle infused $TiO_2$ nanomaterial served the purpose of demonstrating the applicability of the developed method, more advanced high temperature sensing materials are anticipated to offer further opportunities for sensor optimization. As shown, there is a strong potential for the detection of chemical gradients by combining existing techniques such as optical frequency domain reflectometry with the developed nano-engineering enabled optical fiber sensor. In a future publication we hope to demonstrate the measurement of concentration gradients in an actual device, such as a solid oxide fuel cell.

### Acknowledgements

This work was supported by the National Science Foundation (CMMI-1054652, and CMMI-1300273) and the Department of Energy (DE-FE0003859)).

This report was prepared as an account of work sponsored by an agency of the United States Government. Neither the United States Government nor any agency thereof, nor any of their employees, makes any warranty, express or implied, or assumes any legal liability or responsibility for the accuracy, completeness, or usefulness of any information, apparatus, product, or process disclosed, or represents that its use would not infringe privately owned rights. Reference herein to any specific commercial product, process, or service by trade name, trademark, manufacturer, or otherwise does not necessarily constitute or imply its endorsement, recommendation, or favouring by the United States Government or any agency thereof. The views and opinions of authors expressed herein do not necessarily state or reflect those of the United States Government or any agency thereof.